\documentclass[12pt,epsf]{article}
\usepackage{epsfig}
\newlength{\abstractwidth}
\newcommand{\Hilbert}{{\mathcal{H}}}
\newcommand{\Surface}{{\mathcal{S}}}
\newcommand{\Region}{{\mathcal{R}}}
\newcommand{\BB}{{\mathcal{B}}}
\newcommand{\LL}{{\mathcal{L}}}

\newcommand{\TL}{\tilde{\mathcal{L}}}

\newcommand{\starttext}{
\setcounter{footnote}{0}
\renewcommand{\thefootnote}{\arabic{footnote}}}
\newcommand{\be}{\begin{equation}}
\newcommand{\ee}{\end{equation}}

\begin{document}
\renewcommand{\theequation}{\thesection.\arabic{equation}}
\begin{titlepage}
\bigskip
\rightline{SU-ITP 01-20}
\rightline{hep-th/0104111}

\bigskip\bigskip\bigskip\bigskip

\centerline{\Large \bf {Virtual Reality in the World of Holograms}}

\bigskip\bigskip
\bigskip\bigskip

\centerline{\it Michal Fabinger}
\medskip
\medskip
\medskip
\medskip

 \centerline{Department of Physics} \centerline{
Stanford University} \centerline{Stanford, CA 94305-4060}

\medskip
\centerline{\tt fabinger@stanford.edu}
\medskip
\medskip
\bigskip\bigskip
\begin{abstract}

If we assume that the initial conditions for the universe were such that there was no volume-extensive entropy `at the beginning of time' (which is true in Linde's chaotic inflation), we can formulate a covariant holographic bound on the entanglement entropy inside or outside closed space-like surfaces. This bound should hold even for regions where the coarse-grained entropy exceeds the surface area. We find that Bousso's bound gives strong support for this conjecture. We also present a speculative interpretation of the entropy bound, according to which any observer of interest can be  surrounded by a holographic screen, providing a non-redundant description of the rest of the universe. 

\medskip
\noindent
\end{abstract}

\end{titlepage}
\starttext \baselineskip=18pt \setcounter{footnote}{0}

\setcounter{equation}{0}
\section{Introduction}

One of the most intriguing ideas that emerged in theoretical 
physics during the last decade was definitely the holographic principle, 
formulated by 't Hooft  and Susskind \cite{tHooft, Hologram}, 
whose considerations stemmed from earlier work by Bekenstein 
\cite{Bekenstein1, Bekenstein2} and Hawking \cite{Hawking}. According to that 
principle, the logarithm of the number of different possible quantum states 
of a physical system is bounded by its surface area in Planck units, 
which strongly contradicts our expectations based on quantum field theory.
 The precise meaning of this general statement 
was, however, unclear. It is because in various physical situations, there are problems with the entropy, which should be, of course, bounded by the same  number as well.  One of the  most important
examples is the case of  very large regions (much larger than the Hubble scale),
in cosmological models with 
a uniform entropy density, where the volume grows faster than the 
surface area,  and thus at some point the coarse-grained entropy violates 
the holographic bound. 

This motivated Fischler and Susskind  to formulate
their cosmological bound on entropy passing through certain light-like
hypersurfaces \cite{FischlerSusskind}. Their conjecture was  
improved and generalized by Bousso to a form \cite{Bousso} which 
was subsequently proven by Flanagan, Marlof and Wald \cite{FMW}
under two different sets of hypotheses about the entropy flux.
Unfortunately, Bousso's conjecture alone does not seem to have good 
properties for actual construction of a 
general holographic theory \cite{Bousso, TavakolEllis}, and thus, the question of holography in general spacetimes remains open.

In the meantime, string theory went through a period of very rapid progress.
Two of the most important results, Matrix theory \cite{BFSS} and
especially Maldacena's AdS-CFT correspondence \cite{ADSCFT},
can be viewed as realizations of the holographic principle, which 
justifies Susskind's claim \cite{Hologram}
that string theory should be holographic.
This strongly supports the hopes that it will be possible to describe 
gravity without changing the principles of  quantum mechanics.

However, it is far from clear what should be an appropriate description
of our universe within the present-day string theory. Many important
stringy arguments rely on unbroken supersymmetry, which we do not observe
in nature. Non-supersymmetric configurations of string 
theory suffer in general from the cosmological constant problem
\cite{CosmologicalConstant}, and they may also undergo a
catastrophic vacuum decay  \cite{VaccumDecay}. Therefore,
the holographic  principle seems to be one of the very few connections
between strings and the real world.

Even if we had a description of our universe of an AdS-CFT type,
we should not be fully satisfied. The AdS-CFT correspondence
describes the spacetime only from the point of view of observers
at infinity, while we can easily imagine observers falling into
black holes as well. Should we discriminate against those observers, just because
the holes are black? That would certainly be against the principle of
general relativity, which states that it must be possible to describe
nature from the point of view of any observer.

Actually, there is a 
discrepancy between quantum mechanics and general relativity
even on the most fundamental level. According to quantum 
mechanics, only quantities which are in principle observable 
have physical significance, whereas general relativity predicts
the existence of causal horizons, and therefore not everything 
should be accessible to our observations.
An even more striking discrepancy arises when we think about
an evaporating black hole (Fig. \ref{Evaporation}) and
when we assume that its evolution is unitary (at least for observers at infinity).
In that case, we arrive
at a paradox of duplication of quantum information. It seems
that the only possible resolution to it is the black hole 
complementarity principle \cite{BHCP}, which states that
the laws of quantum mechanics are not violated, provided we 
consider only quantities seen by just one particular observer.
Because we are interested in the possibility of having a theory
of quantum gravity, we will adopt this philosophy throughout 
the whole paper.

\vskip 1.5 mm
{\bf Overview}
\vskip 1.5 mm

The main idea of this paper stems from the fact that in most inflationary cosmological models, an overwhelming majority of the coarse-grained entropy was created after inflation or at the end of it. Therefore, we do not need any volume-extensive entropy to be present in the universe before inflation in order to be consistent with observations. If we now use Occam's razor, we can even say that `at the beginning of time,' there was no volume-extensive entropy at all. (Actually, this is a natural assumption of Linde's chaotic inflation \cite{ChaoticInflation}.) If this is really true, it has interesting consequences: in this case, we can conjecture that there is a `holographic' bound on the entanglement entropy inside (or outside) closed space-like surfaces. These surfaces should be observable ({\it i.e.} there must be a spacetime-point in the causal future of the whole surface), and they should satisfy the generalized convexity condition of Section \ref{ConditionOnHolographicSurfaces}, by which we mean a certain condition of the Bousso type (saying that the expansion of any light-like hypersurface orthogonal to the surface of interest should not change its sign as we go from one point of the surface to another). The  conjecture itself then states, roughly speaking, that the entanglement entropy inside (or outside) any region bounded by such a surface is not larger than the surface area $A$ (in Planck units). Here, the entropy can be the entanglement entropy with respect to any other physical system.

Presumably, there are two sources of the entanglement entropy: entanglement in the vacuum and the presence of `real particles', which can be possibly entangled. It is impossible to clearly distinguish these two contributions, but it will not be crucial for our further discussion.
We can expect that the entanglement entropy due to quantum fluctuations in the vacuum should not exceed the surface area (see Section \ref{EntanglementInTheVacuum}), but we are unable to make any more precise statement about it. On the other hand, the contribution of `real particles' should be nicely bounded by $A/4$, as we can argue using Bousso's bound \cite{Bousso, FMW}.

In addition, we present a very radical interpretation of this bound (which can be wrong even if the conjecture itself is correct). According to it, the surface area bounds not only the entanglement entropy, but also the logarithm of the dimension of the Hilbert space which 
is capable of describing all the possible quantum states of the respective 
region.
This is in spite of the fact that the coarse-grained entropy can be many orders of magnitude larger than $A$. We argue that the coarse-grained entropy  is not relevant for the dimension of the true Hilbert space in quantum gravity, because it ignores subtle quantum correlations between the state of matter in cosmologically large regions, which must be (and actually are) according to inflationary cosmology really present. 

Finally, we mention a possible weaker form of the conjecture, where we restrict ourselves just to those surfaces for which there exists at least one event in the causal past of the whole surface. The validity of such a  formulation of the conjecture does not depend on any particular cosmological assumption. 

\section{ Entropy and its properties }

In this section we will review some properties of entropy which we will need in the rest of the paper.

\subsection{ Entropy, coarse-grained entropy, and entanglement entropy }
The {\it fine-grained entropy} ({\it von Neumann entropy}\ or simply {\it entropy}) of a quantum mechanical system which is  in a mixed state described by a density matrix $\rho$ is defined 
as\footnote{ Strictly speaking, this is the definition of entropy in non-relativistic quantum mechanics. In quantum field theory, we must also specify a regularization scheme, and to make this quantity finite, we have to discard the contribution of quantum fluctuations, which is for localized systems nonzero and ultraviolet divergent (as discussed in Section \ref{EntanglementInTheVacuum}). However, we expect that this divergence will be resolved in the theory of quantum gravity.
Also note that in this paper we will use extensively the Planck units $\hbar = c = G = k_B = 1$.}
\be
S = - \hbox{Tr}(\rho \ln \rho).
\label{Entropy}
\ee
If the density matrix $\rho$ is thermal, then this definition coincides with what is called `entropy' in phenomenological thermodynamics. However, when the system is in a pure state  $|\psi \rangle$, the definition  (\ref{Entropy}) gives us zero, in spite of the fact that the pure state can  be without any problems a typical representative of a thermal ensemble. This contradicts our
thermodynamical intuition, and that is why we usually introduce the  procedure called {\it coarse-graining}. One of the ways to think about coarse-graining is as follows: We divide the whole system with Hilbert space $\Hilbert$ into a lot of local subsystems with Hilbert spaces $\Hilbert_i$,
\be
\Hilbert = \Hilbert_1  \otimes  \Hilbert_2  \otimes  ... \  \Hilbert_{n}.
\label{Factorization}
\ee
Then, we compute the reduced density matrices 
\be 
\rho_i = \hbox{Tr}_{\Hilbert_1, \Hilbert_2 ... \Hilbert_{n} \ {\rm except}  \  \Hilbert_i}
              |\psi \rangle \langle \psi |
\ee
and the corresponding entropies
\be
S_i = - \hbox{Tr}_{\Hilbert_i} (\rho_i \ln \rho_i).
\label{SubsystemEntropy}
\ee
When we add these entropies together,
\be
S^{\rm cg} = \sum_i S_i,
\label{CoarseGrainedEntropy}
\ee
we obtain the desired 
answer,\footnote{Here, we have to discard the contribution of quantum fluctuations (discussed in Section \ref{EntanglementInTheVacuum}) even if we believe that the ultraviolet divergence will be resolved in the theory of quantum gravity. This is because the entropy due to quantum fluctuations is proportional to the surface area of the physical system, and thus for a large number of small systems, we would get a quantity without any physical meaning.}
which we call {\it coarse-grained entropy}. It is this kind of entropy the second law talks about.
Note that the coarse-grained entropy (\ref{CoarseGrainedEntropy}) can never be smaller than the
fine-grained entropy (\ref{Entropy}),
\be
S \leq S^{\rm cg}.
\label{SLessThanSCG}
\ee
This is because of the triangle inequality \cite{ArakiLieb}
\be
| S_1 - S_2 | \leq S \leq S_1 + S_2,
\label{Triangle}
\ee
which holds in the case of dividing a system into two subsystems.

Let us  focus  on the case of two subsystems a little bit more. 
From (\ref{Triangle}) it follows that if the overall system is in a pure state, $\rho = | \psi \rangle \langle \psi |$, the entropy $S_1$ must be equal to $S_2$. We refer to it as  {\it entanglement entropy}. Now, suppose that the first subsystem is a physical system we want to study and the second one corresponds to its surroundings, which include some measuring apparatus. Also suppose that the overall system is completely isolated and that it is in a pure 
state.\footnote{This is without loss of generality, because a mixed state can be thought of as an ensemble of pure states with different probabilities, and we will not consider performing a measurement on the overall system with an external apparatus.}
 After a measurement of the system of interest ({\it i.e.} of the first subsystem), it will be in general entangled with its surroundings ({\it i.e.} the apparatus and other parts of the second subsystem, for example some particles escaping to infinity). If there was any limitation on the maximum degree of entanglement we could produce, there would also be a limitation on the number of states we could distinguish with our apparatus. With an ideal apparatus, we would be able to distinguish all the states of the studied system. In that case, the maximum entanglement entropy which could be in principle produced by the measurement would coincide with the logarithm of the dimension of $\Hilbert_1$.

\subsection{ Troubles with coarse-grained entropy in general relativity} 
\label{TroublesWithCoarseGrainedEntropyInGeneralRelativity}

\begin{figure}[htbp]
\centerline{\epsfysize=1.80truein \epsfbox{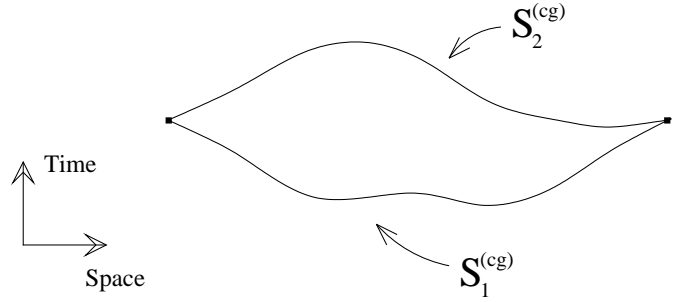}} \vskip -0.4
cm
\caption{\small \sl This picture shows two time-slices (having the same boundary)  through the volume of a certain spacetime region. According to the second law, the coarse-grained entropy $S_1^{\rm cg}$ 
cannot be larger than $S_2^{\rm cg}$, but it can definitely be lower. On the other hand, because of unitarity of the evolution,  the fine-grained entropy is the same in both cases. It is independent of time-slicing, and therefore, it might be well defined even in those cases when the inside geometry is not a reasonable quantity.
}  
\label{Slices}
\end{figure}

In order to compute the coarse-grained entropy of a certain region of spacetime, we always have to specify the time-slice on which the entropy should live. Specifying only the boundary is not sufficient, since the coarse-grained entropy can grow in time, and in the situation on 
Fig.\ref{Slices}, $S_2^{\rm cg}$ can be greater then $S_1^{\rm cg}$. It certainly cannot be less then $S_1^{\rm cg}$, because we expect the second law of thermodynamics to hold in this situation. 
\begin{figure}[htbp]
\centerline{\epsfysize=3.00truein \epsfbox{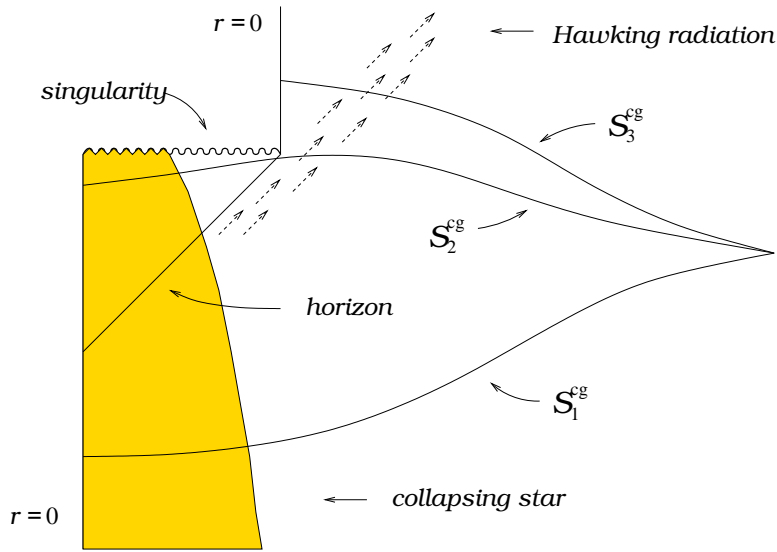}} \vskip -0.4
cm
\caption{\small \sl In the case of an evaporating black hole, coarse-grained entropy loses its physical meaning. The second time slice would contain the same entropy twice, once in the form of the matter of the collapsing star and once in the form of the Hawking radiation.}
\label{Evaporation}
\end{figure}
This works well, but for example in the case of an evaporating black hole \cite{Hawking}, we encounter serious problems with coarse-grained entropy. Fig. \ref{Evaporation} shows a collapsing star forming a black hole, which finally evaporates. On the first time-slice, the coarse-grained entropy $S_1^{\rm cg}$ is just the entropy of the collapsing star. On the second time-slice, the coarse-grained entropy $S_2^{\rm cg}$ corresponds to the entropy of the collapsing star plus the entropy of the Hawking radiation. Finally, the coarse-grained entropy $S_3^{\rm cg}$ on the third time-slice is the entropy of the Hawking radiation only. $S_2^{\rm cg}$ can be easily greater than 
$S_3^{\rm cg}$, so a simple application of the second law is impossible in this case.
Clearly, there is an overcounting on the second time-slice. The same entropy contributes to $S_2^{\rm cg}$ twice - once in the form of the matter of the star and once in the form of the Hawking radiation.  If we want some kind of second law to be valid (for an observer at infinity), we must not think about what is happening inside the horizon, but instead, we should treat the whole black hole as one object. This is actually the philosophy of the generalized second law \cite{Bekenstein1}. 

In fact, there is an even more serious problem with this example, which we mentioned already in the Introduction. It arises when we assume that the evolution is unitary for the observer at infinity. (And we will do so in this paper.)  Not only does the coarse-grained entropy (\ref{CoarseGrainedEntropy}) not have much physical significance here, but even the factorization of the Hilbert space $\Hilbert$ describing the situation from the point of view of an observer at infinity into local Hilbert spaces (\ref{Factorization}) is impossible. If we thought that on the second time-slice in Fig. \ref{Evaporation}, the quantum information can be objectively localized, we would have to conclude that the quantum state of the collapsing star is duplicated. It should be there once in the form of the matter of the star, and it should also be encoded in the state of the Hawking radiation. This would definitely violate linearity of quantum mechanics \cite{BHCP}. However, the black hole complementarity principle \cite{BHCP} can solve this apparent paradox very nicely. According to the considerations in \cite{BHCP}, no single observer can ever see such a duplication, and the fundamental laws of quantum mechanics are perfectly valid from the point of view of any of them. Therefore, only quantities which can be actually measured by an observer have any physical significance. We will return to the discussion of the subjectivity of physics also in Section \ref{VirtualReality}.

\subsection{Bousso's holographic bound}
\label{BoussosHolographicBound}

Since we will need Bousso's holographic bound \cite{Bousso} in Sections \ref{EntropyInInflationaryCosmology} and \ref{SupportForTheConjecture}, it will be convenient to describe it shortly here. 

\begin{figure}[htbp]
\centerline{\epsfysize=2.00truein \epsfbox{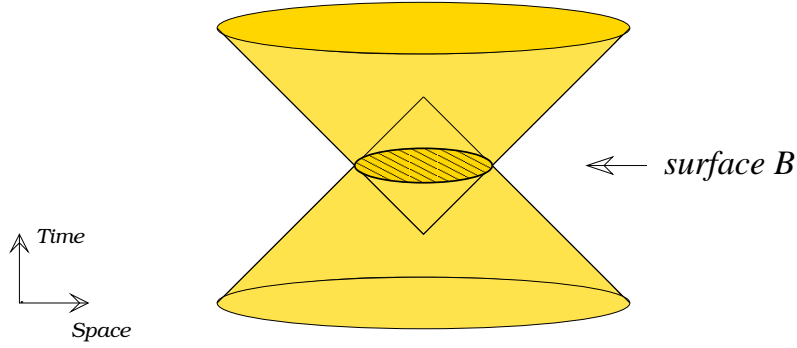}} \vskip -0.4
cm
\caption{\small \sl This picture shows four null hypersurfaces orthogonal to a spherical surface $B$ in Minkowski spacetime, with one spatial dimension suppressed. 
The two hypersurfaces which contain the tips of the cones have a non-positive expansion on the whole surface $B$, and therefore they are light-sheets in Bousso's terminology.
} 
\label{Cones}
\end{figure}

Bousso's conjecture \cite{Bousso} applies to any smooth open or closed space-like surface in any space-time with reasonable energy conditions. First, we have to construct four null hypersurfaces spanned by a congruence of null geodesics which are orthogonal to the surface. (See Fig. \ref{Cones}.)

\begin{figure}[htbp]
\centerline{\epsfysize=1.30truein \epsfbox{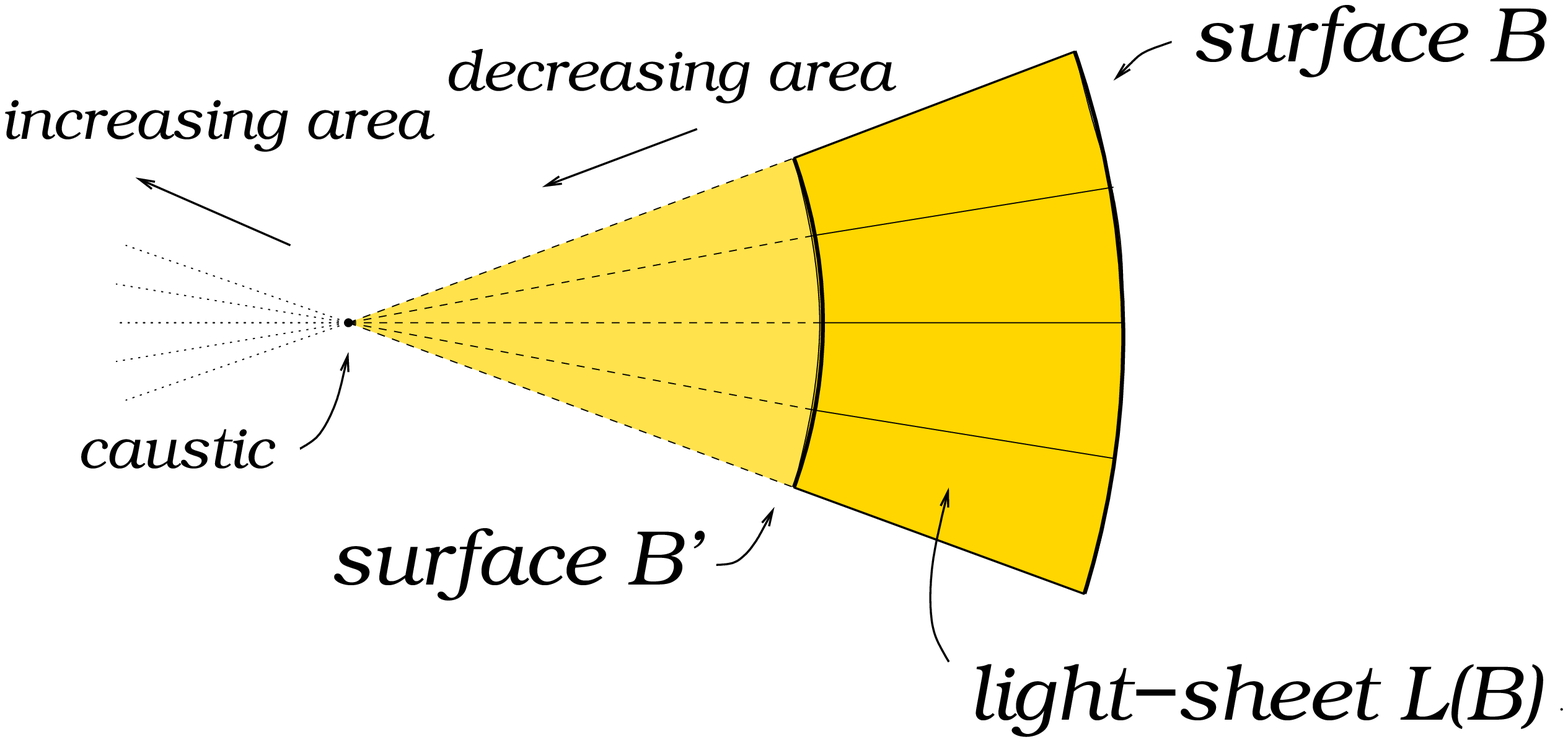}} \vskip -0.4 cm
\caption{\small \sl The expansion $\theta$ is the relative change of a small cross-sectional area of a hypersurface, when we follow the light-rays. Once it is non-positive, it cannot increase any more before reaching a caustic.}
\label{Expansion}
\end{figure}

Then, at every point of the surface and for each hypersurface, we have to calculate the {\it expansion} $\theta$, which 
is defined as 
\be
\theta = \nabla_{\mu}  k^{\mu},
\ee
where $k^{\mu}$ are tangent vectors of the light-rays. 
This quantity
measures how much a small cross-sectional area $\delta A$ changes when we follow the light rays a little bit (see Fig. \ref{Expansion}),
\be
\theta = {1 \over \delta A} \  {d  \over d \lambda} \ \delta A.
\ee

Here, $\lambda$ denotes the affine parameter of the light-ray congruence. 
It is chosen to increase in the direction from the surface.
Now, we have to split the original surface into pieces, in which the $\theta$s do not change their sign as we go from one point to another. Let us focus on one such piece, which we will denote $\BB$. At least two of the corresponding hypersurfaces will have initially a non-positive expansion,
\be
\theta \leq 0.
\ee
Due to the focusing theorem of general relativity \cite{MTW}, $d\theta/ d\lambda$ must be non-positive, and therefore the expansion cannot increase before we reach possible caustics 
({\it i.e.} points where $\theta \rightarrow - \infty$). When we truncate the hypersurfaces at the caustics, we obtain Bousso's light-sheets $\LL (\BB)$. Bousso's conjecture then tells us that the coarse-grained entropy passing through any light-sheet $\LL (\BB)$ is bounded by one quarter of the area of $\BB$,
\be
S^{\rm cg} \leq {A \over 4}.
\ee

In paper \cite{FMW}, a convenient generalization of Bousso's bound was formulated. It states that if we terminate the light-sheets at any (connected or disconnected) space-like surface $\BB'$  of area $A'$ before reaching singularities or caustics (see Fig. \ref{Expansion}), we can strengthen the bound to
\be
S^{\rm cg} \leq {A - A'  \over 4}.
\ee

\section{Motivation for the conjecture}

\subsection{Entanglement in the vacuum }
\label{EntanglementInTheVacuum}

When the fields of a quantum field theory are in some vacuum state  and when we divide the  whole space into two parts, the fields in each part separately will be in a mixed state. 
Then we can ask how large the corresponding entanglement entropy $S$ is. Although we do not know the answer to this question for a general case, in some examples  it was shown that the leading term will be proportional to the surface area $A$ of the boundary of the two regions  \cite{EE}.
\be
S \sim M_{\rm _{UV}}^2 A
\label{SMA}
\ee
Here, $M_{\rm  _{UV}}$ represents the ultraviolet cut-off in the field theory, and we implicitly assume three spatial dimensions. The entropy (\ref{SMA}) is infrared finite (as opposed to the case of just one spatial dimension \cite{OneD}), but clearly, it is ultraviolet divergent. If quantum field theory was a correct description of our world at all energies, it would  be really infinite. However, we expect that at least at the Planck scale, the field theory description breaks down. Therefore, the field-theoretic contribution to the entanglement entropy should be at most of order $A$ (since the Planck mass is equal to one in our units),
\be
S \leq A.
\label{SA}
\ee
Note that if the quantum field theory breaks down at some lower scale than the Planck scale, this inequality can only become sharper. 
Because the main field-theoretic contribution to the entanglement entropy comes from short-wavelength modes near the surface, it should not depend much on the spacetime curvature. For this reason we can assume in the rest of the paper that (\ref{SA}) is valid in general at least for surfaces satisfying the generalized convexity condition of Section \ref{ConditionOnHolographicSurfaces}. The question is, of course, if considering also the physics above the scale where the field theory breaks down could change the approximate inequality (\ref{SA}). However, it is natural to expect that it can not, because otherwise the usual black hole entropy $A/4$ would seem to be less than the entropy of its thermal atmosphere.

\subsection{Entropy in Linde's theory of chaotic inflation}
\label{EntropyInInflationaryCosmology}

Inflationary cosmology \cite{Inflation} is now a standard cosmological scenario which is used to explain the flatness and the large scale homogeneity of our universe and which successfuly passed several observational tests \cite{TestsOfInflation}.
It assumes that just after the `beginning of time', there was an extremely rapid expansion of the universe, called inflation. 

After the earliest papers on inflation \cite{EarliestPapers},
a very nice and simple cosmological model was proposed \cite{Guth}, which is now referred to as the old inflationary scenario. It assumed that the universe was initially very hot, which led to symmetry restoration, and after that, when it cooled down sufficiently, the symmetric state no longer had the lowest possible energy and therefore started to decay. However, as pointed out already in \cite{Guth}, this model produced unacceptably large density perturbations. This was resolved in the new inflationary scenario \cite{NewInflation} (also assuming a hot early universe), but this suffered from its own difficulties. A short time after it was proposed, it was realized that in such a scenario, the theory of high temperature symmetry restoration was not applicable, and that inflation would start so late that the universe could easily collapse or become too inhomogeneous during such a long time. Also, it would be necessary to assume homogeneity of the corresponding region from the early beginning. The modern inflationary models \cite{ChaoticInflation, Inflation} which  are called `chaotic inflation' in a generalized sense have the beautiful property that they can assume that inflation started at the Planck time and that all parts of our universe originated from a region of the Planck size (which is Linde's assumption). Possible alternatives to this assumption are to expect that inflation started later or in a larger region and to use for example the anthropic principle to explain why it actually happened, or to expect some exotic physics at sub-Planckian energies. The assumption itself is however very attractive, and therefore it might be interesting to discuss its consequences for the fine-grained entropy in our universe:

If our universe really originates from a region of the Planck size, then we expect that at the beginning, there were no particles in the universe at all. The coarse-grained entropy we observe now was then produced during the evolution of the universe, mostly during and after the so called
reheating\footnote{Originally, reheating used to be modeled only as a perturbative decay of the inflaton field into different particles. However, the temperatures produced in such models cannot be very high, and that is why cosmologists consider now a more efficient non-linear process, referred to as `preheating' \cite{Preheating}.}
period at the end of inflation, when particles were created at the cost of the energy of the inflaton field (a scalar field, or maybe fields, driving the dynamics of inflation). 

\begin{figure}[htbp]
\centerline{\epsfysize=2.40truein \epsfbox{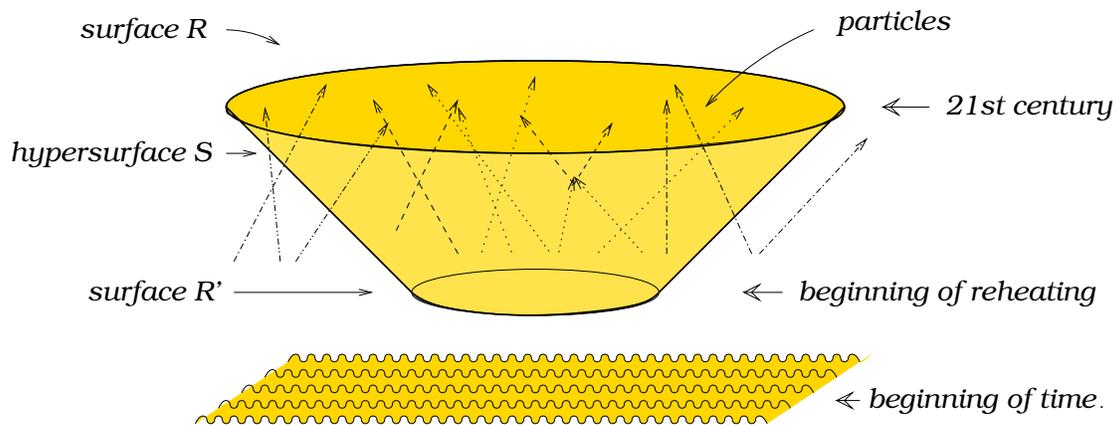}} \vskip 0.4
cm
\caption{\small \sl According to Linde's chaotic inflation, all the coarse-grained entropy of `real particles' we observe now was produced during and after the reheating period at the end of inflation. 
For this reason and because of the unitarity of the evolution, there are subtle quantum correlations between the state of matter in  different parts of cosmologically large 
regions.
They cause the fine-grained entropy of region $\Region$ in this picture to be less than the corresponding coarse-grained entropy. In fact, the fine-grained entropy inside $\Region$ is bounded roughly by its surface area, independently of its size. (Note however, that this picture is absolutely not to scale.)}
\label{Inflation}
\end{figure}

Let us now think about a spherical region $\Region$ larger than the observable universe (by which we mean the region from which we can get any information in the form of particles, {\it i.e.} a region roughly of the size of the inverse Hubble constant). As is clear from Fig. \ref{Inflation}, the ({\it fine-grained}) entropy $S_{\Region}$ of the region $\Region$ is certainly less than or equal to the sum of the entropy $S_{\Region'}$ of the region $\Region'$ and the entropy $S_{\Surface}$ that passed through the light-like hypersurface $\Surface$,
\be
S_{\Region} \leq S_{\Region'} + S_{\Surface}.
\ee
(See also Fig. \ref{Slices}.)
For  the 
`semiclassical part'\footnote{
The distinction between the `semiclassical' entropy of `real particles' and the entropy due to quantum fluctuations is merely our choice of point of view. However, it is often convenient to make it.
}
 of $S_{\Surface}$, we can use Bousso's bound \cite{Bousso}, which here coincides with the Fischler--Susskind bound \cite{FischlerSusskind}. Therefore, the semiclassical part of $S_{\Surface}$ must be bounded by $A_{\Region}/4$. (Strictly speaking, Bousso's conjecture talks only about the coarse-grained entropy passing through the light-like hypersurface $\Surface$, but it obviously cannot be smaller than the contribution we are interested in.) Because $A_{\Region'} \ll A_{\Region}$, we can also say that the contribution of quantum fluctuations is presumably at most of order of $A_{\Region}$ (according to Section \ref{EntanglementInTheVacuum}). Putting these two estimates together,
\be
S_{\Surface} \leq A_{\Region}.
\label{SSurface}
\ee
The entropy $S_{\Region'}$ does not have any `semiclassical' part at all, because at the beginning of reheating, all the entropy was just in the form of quantum fluctuations of fields.  According to Section \ref{EntanglementInTheVacuum}, we can assume that it was not larger than of order $A_{\Region'}$, which is much less than $A_{\Region}$,
\be
S_{\Region'} \leq A_{\Region'} \ll A_{\Region}.
\label{SRegion'}
\ee
(If we decided to extend the light-like hypersurface $\Surface$ up to the big bang singularity, the corresponding entropy $S_{\Region'}$ would essentially vanish, since as we 
said, in this section, we are interested in the consequences of the whole universe stemming  from a region of the Planck size. This way of expressing the idea is somewhat simpler, and that is why we will use a similar one in Section \ref{SupportForTheConjecture}) Now, we can combine
(\ref{SSurface}) and (\ref{SRegion'}) to obtain
\be
S_{\Region} \leq A_{\Region}.
\label{SLessThanA}
\ee
For extremely large regions $\Region$, the entropy $S_{\Region}$ could be many orders of magnitude less than the corresponding coarse-grained entropy, because of subtle quantum correlations between the states of the particles produced during reheating\footnote{The significance of quantum correlations between particles in
cosmologically large regions for the holographic principle was discussed in \cite{Smolin}.}. From the point of view of the holographic principle, we should not be interested in the maximum possible coarse-grained entropy, but in the maximum fine-grained entropy. The result (\ref{SLessThanA}) is therefore very encouraging.

\subsection{The Heisenberg story}

The holographic principle, which we would like to discuss in this paper, is roughly speaking the  statement that quantum field theory overestimates the number of possible states in nature and that in a theory of quantum gravity, this number should be much lower. In fact, such a radical reduction of possible states already appeared in the history of physics in the transition from classical to quantum mechanics, and it might be interesting to discuss it a little bit closer in an example.

Consider a microcanonical ensemble for a dilute gas which can be described very well within the framework of classical mechanics. We could ask what the number  of different microstates for a given macroscopic state is. If we took the formulation of this question seriously, we would conclude that  it is infinity, because there is an infinite number of points in the phase space which correspond to the same values of macroscopic observables. This is why in classical mechanics, entropy can be defined only up to an additive constant. On the other hand in quantum mechanics, where we have to take Heisenberg's uncertainity relations into account, we find that 
although the number of accessible states  even for a narrow energy interval is huge, it is definitely finite. For this reason, entropy in quantum mechanics can be defined without using an arbitrary additive constant. 

 We see that it is possible to have a physical system within the range of validity of an approximate description, and to calculate correctly almost all its properties within that framework, but when the number of different states comes into question, the approximate description can give us a wrong answer.

\subsection{Quantum states in quantum gravity}
\label{QuantumStatesInQuantumGravity}

Most of the discussion in this section is relevant just for the speculative interpretation of the entropy conjecture. However, it is one of the main motivations for proposing it.

We have seen in Section \ref{TroublesWithCoarseGrainedEntropyInGeneralRelativity} that when strong gravitational effects are taken into account, it does not always make sense to localize quantum information in the volume of spacetime regions, because some part of it can be hidden behind a causal horizon and thus irrelevant for the observer we are interested in. However, there is one more possibility. We can associate quantum states with the boundaries of the regions. If the boundaries themselves are accessible to the observer, we should not arrive at any paradox. In fact, a very similar approach was adopted already in the considerations related the black hole complementarity principle (mentioned in Section \ref{TroublesWithCoarseGrainedEntropyInGeneralRelativity}), when the concept of `stretched horizon' was introduced \cite{BHCP}.

To respect this principle, we will always be interested only in those regions which are `{\it observable}'. By this we will mean regions for which there exists at least one spacetime point in the causal future of their whole boundary. When we talk about `systems' or `regions', we will always assume that they are observable in the above sense, unless otherwise noted.

\begin{figure}[htbp]
\centerline{\epsfysize=2.00truein \epsfbox{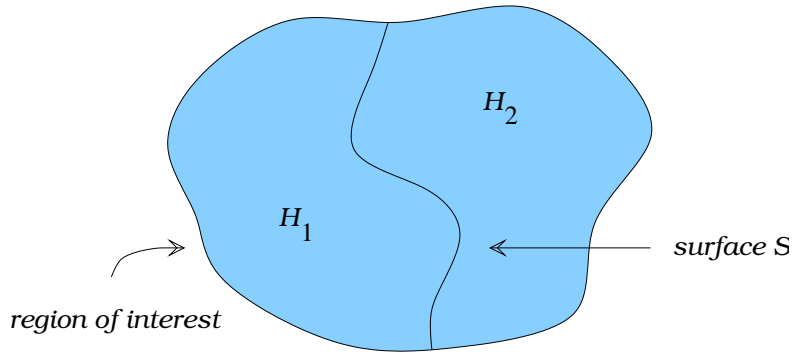}} \vskip 0 cm
\caption{\small \sl Whenever a spacetime region  contains a suitable surface such as $\Surface$ in this picture, we can use it to cut the region into two pieces (if they are both observable). Each piece should be described by some Hilbert space, but their product does not necessarily need to be  contained in the Hilbert space of the whole region. Such a possibility would contradict our expectation based on quantum field theory, but not the fundamental principles of quantum mechanics.}
\label{Cut}
\end{figure}

Let us be a little bit more concrete about associating quantum states with the boundaries of regions. With each boundary geometry (and other boundary conditions), we would like to associate a Hilbert space $\Hilbert$, describing the quantum states that can be inside the boundary. Presumably, some vectors in this Hilbert space could correspond to states with different inside geometries, or to linear superpositions of them, whenever the term `inside geometry' makes sense. In those cases, when we can use the semiclassical approximation for the geometry, there should be just one inside geometry which dominates. Consider now a quantum state that 
corresponds to a definite geometry of a certain surface $\Surface$ which can be used to cut the region into two pieces. This situation is depicted in Fig. \ref{Cut}. 
The possible quantum states inside the  pieces should be described by some Hilbert spaces $\Hilbert_1$ and $\Hilbert_2$. Note however, that the product $\Hilbert_1 \otimes \Hilbert_2 $ does not necessarily have to be contained in $\Hilbert$, although our intuition tells us something different. This is just intuition based on quantum field theory, and presumably, it can be wrong when both gravity and quantum theory must be taken into account. We have seen in the example of an evaporating black hole how crazy quantum theory can be when it enters the realm of general relativity. In fact, there is really no quantum-mechanical principle requiring $\Hilbert_1 \otimes \Hilbert_2 \subseteq \Hilbert $ (and therefore
the maximum entropy does not necessarily have to be an extensive quantity). The only thing we need is the possibility to write any state $|\psi\rangle_{\Hilbert} \in \Hilbert$ which `contains' the given surface $\Surface$ as
\be
|\psi\rangle_{\Hilbert}  = \sum_{ij} c_{ij} |i\rangle_{\Hilbert_1} |j\rangle_{\Hilbert_2}, 
 \ee
where $|i\rangle_{\Hilbert_1}$ and $ |j\rangle_{\Hilbert_2}$ form the bases of $\Hilbert_1$ and $\Hilbert_2$. 

It is important to note that it does not make any sense to define some Hilbert space for a region whose boundary is not fixed. If we change the geometry of spacetime, then its absolutely unclear what still belongs into the region of interest and what does not. Therefore, the only sensible Hilbert spaces in curved spacetime might be those which would describe possible quantum states inside some definite boundary geometry (maybe also with some other boundary conditions). If we want to talk about the evolution of a region whose boundary is not fixed, we should use different Hilbert spaces for different time-slices of the boundary.

Also, it seems to be inconsistent with the original meaning of quantum states to assign some Hilbert space to the whole universe, which would not exclude any region where the observer should be. If we had to do so, then the original principles of quantum mechanics would have to be modified. In this paper, we do not expect that such a modification is necessary. The `Hilbert space of the whole universe' should be trivial.


\section{The conjecture}

\subsection{Condition on holographic surfaces}
\label{ConditionOnHolographicSurfaces}

Motivated by our previous discussion, we will attempt to formulate a holographic bound on the entropy inside closed space-like surfaces. However, there is one obstruction pointed out in \cite{FMW}. If we take any two-surface that encloses a given world-tube, we can always find another two-surface which is arbitrarily close to the original surface, encloses the same world-tube and which has an arbitrarily small area. (It can be obtained by `wiggling' the original surface suitably in spacetime.) To overcome this difficulty, we have to impose an appropriate condition on the holographic surfaces. Since we want our bound to be covariant, it is natural to impose the following condition of the Bousso type.    

{\bf Generalized convexity condition.} {\it For any given closed space-like 
two-surface\footnote{
Here, as well as in the other parts of the paper, we implicitly assume that the surface is smooth.
},
$\BB$ we construct null hypersurfaces $\LL_1 (\BB), \LL_2 (\BB), \LL_3 (\BB)$ and  $\LL_4 (\BB)$ orthogonal to this two-surface. For each hypersurface $\LL_i (\BB)$, we calculate the 
expansion $\theta_i$ at each point of the two-surface. Now, we require that $\theta_i$ does not flip its sign as we go from one point of $\BB$ to another. In other words, for each particular $\LL_i (\BB)$,
we require
$\theta_i \geq 0$ at all points of $\BB$
or
$\theta_i \leq 0 $ at all points of $\BB$.
} 

If this is satisfied, we will consider this two-surface $\BB$ in the entropy bound we will formulate.
Note that this condition is defined locally in the neighborhood of $\BB$ and does not involve quantities deep inside or far outside of $\BB$, as well as quantities in the distant past or future. Also, note that this requirement automatically excludes the problematic surfaces with almost vanishing area.
For surfaces in Minkowski space lying in hypersurfaces of constant Minkowski time, the condition above is just the requirement that the region inside $\BB$ should be convex. 

The conjecture we will formulate is side-symmetric, and therefore there is no need to distinguish `inside' from `outside' in a general spacetime. Only for convenience, when we talk about some specific region, will we refer to it as `inside' of its boundary, with the rest of the space `outside.'

\subsection{Formulation}

\hskip 6mm {\bf Assumption.} {\it Let us assume that the initial conditions for the universe were such that there was no volume-extensive entropy at all at the `beginning of time.' If this is true, the following conjecture should be correct.}

{\bf Conjecture.} {\it Consider any two 
observable\footnote{ In the sense of Section \ref{QuantumStatesInQuantumGravity}.}
subsystems forming together an observable system. Suppose that the boundary of the first subsystem has area $A$ and satisfies the generalized convexity condition of  Section \ref{ConditionOnHolographicSurfaces}. Then the entanglement entropy of the first subsystem with respect to the second subsystem  cannot be  larger than of order $A$.
 }

\subsection{Support for the conjecture}
\label{SupportForTheConjecture}

The entanglement entropy we were talking about should presumably come from two 
contributions\footnote{
As said in a footnote in Section \ref{EntropyInInflationaryCosmology}, the distinction is only our choice of point of view.}.
One of them is the entropy of quantum fluctuations in the `vacuum'. Even though it is difficult to make any precise statement about it, according to our discussion in Section \ref{EntanglementInTheVacuum} we expect this contribution to be at most of order the surface area. About the other contribution, entropy due to `real particles', we can say much more, because whenever the semiclassical approximation is appropriate, we can apply Bousso's bound \cite{Bousso, FMW}, as we did in Section \ref{EntropyInInflationaryCosmology}. 

Let us then discuss the semiclassical entropy of `real particles':

For any surface $\BB$, with area $A$, satisfying the generalized convexity condition of Section \ref{ConditionOnHolographicSurfaces}, we can construct four null hypersurfaces  $\LL_1 (\BB), \LL_2 (\BB), \LL_ 3 (\BB)$ and $\LL_4 (\BB)$ 
orthogonal to this surface and terminated at caustics or singularities. At least two of them have a non-positive expansion $\theta_j$, so they are light-sheets in Bousso's terminology. Then, we truncate these light-sheets whenever non-neighboring light-rays intersect (as in \cite{TavakolEllis}), and also whenever they reach a black hole horizon. In this way, we obtain at least two truncated light-sheets $\TL _j (\BB)$.
Now, at least one of the following possibilities must apply. (`Inside' here corresponds to the first subsystem.)

\vskip 2mm
$\bullet$ {At least one truncated light-sheet $\TL _k (\BB)$ is simply connected, past directed, and ingoing.}
\vskip 2mm

\noindent In this case, we can think of $\TL _k (\BB)$ as the limit of a certain series of time slices through the inside of $\BB$. The coarse-grained entropy on this limiting time-slice is according to Bousso less than or equal to $A/4$. Since the entropy (\ref{Entropy})
can never exceed the coarse-grained entropy (\ref{SLessThanSCG}), the contribution of `real particles' to the fine-grained entropy must be bounded by $A/4$ as well.
Putting the estimates for the two contributions (of `real particles' and quantum fluctuations) together, we see that the entanglement entropy of the region inside $\BB$ with respect to any other physical system cannot be larger than of order $A$.

\vskip 2mm
$\bullet$ {At least one truncated light-sheet $\TL _k (\BB)$ is simply connected, past directed, and outgoing.}
\vskip 2mm

\noindent We can use the reasoning of the previous case to show that the entanglement entropy with respect to the whole outside is bounded by $A$. It is then natural to expect that the entanglement entropy with respect to only some part of the outside cannot be larger than this bound.
\vskip 2mm
$\bullet$ {At least one truncated light-sheet $\TL _k (\BB)$ is non-simply connected, past directed, and ingoing.}
\vskip 2mm

\begin{figure}[htbp]
\centerline{\epsfysize=2.20truein \epsfbox{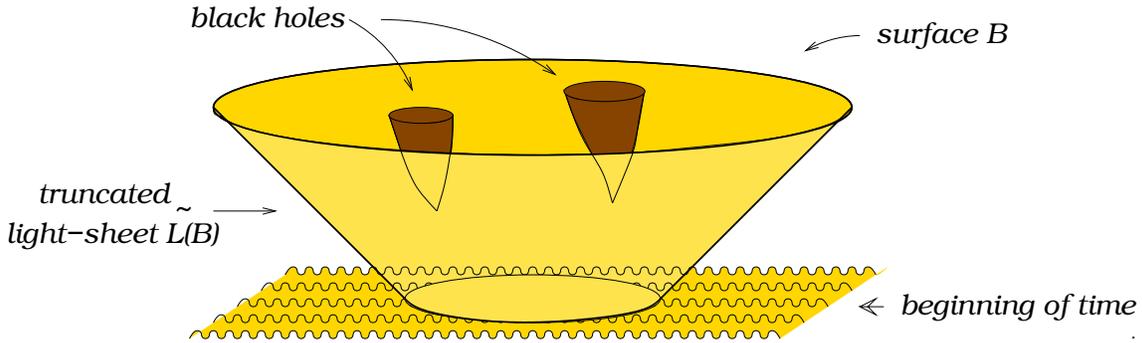}} \vskip 0
cm
\caption{\small \sl The surface $\BB$ in this figure possesses an ingoing past-directed truncated light-sheet which is not simply connected and which therefore extends up to the big bang singularity. We can imagine that the truncated light-sheet $\TL (\BB)$ and the `circle' lying on the big bang singularity form together the limit of a certain series of time-slices through the inside of $\BB$.}
\label{BigBang}
\end{figure}

\noindent The fact that $\TL _k(\BB)$ is not simply connected means that it is terminated at the big bang singularity\footnote{As mentioned in Section \ref{EntropyInInflationaryCosmology}, this formulation is just a shorthand for saying that we are free to truncate the light-sheet at sufficiently early times, when the universe was `empty.'}, as in Fig. \ref{BigBang}. (This is because we do not expect any white holes in our universe. Some time reversed black hole evaporation would be an extremely improbable event.) We can construct a series of time-slices through the inside of $\BB$, whose limit is $\TL _k(\BB)$ plus a certain part of the big bang singularity. Using Bousso's bound, we see that the coarse-grained entropy on $\TL _k(\BB)$ is bounded by $A/4$. The contribution of the part of the limiting time-slice which lies on the big bang singularity is, according to the assumption of this paper, zero. (Remember that here, we were discussing only the entropy of what we call `real particles'. We {\it assume} the entropy of quantum fluctuations to be less than $A$.) Thus, we see that the entanglement entropy of the region inside $\BB$ with respect to any other region is not larger than of order $A$.

\vskip 2mm
$\bullet$ {At least one truncated light-sheet $\TL _k (\BB)$ is non-simply connected, past directed, and outgoing.}
\vskip 2mm

\noindent We can use the argument of the previous case to show that the entanglement entropy with respect to the whole outside is bounded by $A$. Again, we can expect that the entanglement entropy with respect to only some part of the outside cannot be larger than of order $A$.

\begin{figure}[htbp]
\centerline{\epsfysize=2.8truein \epsfbox{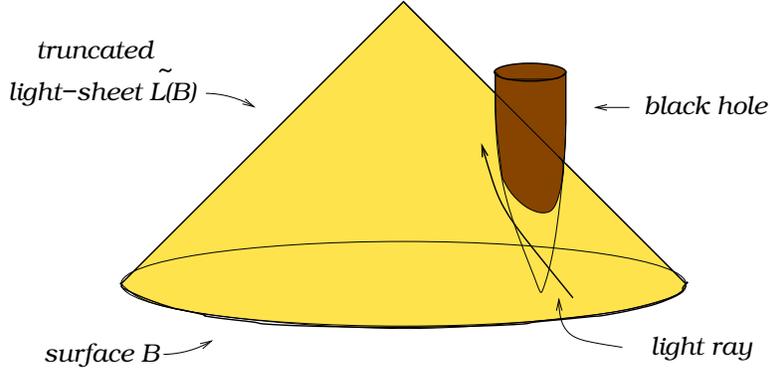}} \vskip 0
cm
\caption{\small \sl This idealized schematic picture shows a region whose boundary has a future directed truncated light-sheet punctured only by a black hole. In this case, we can use the generalized Bousso's bound \cite{FMW}.}
\label{Chimney}
\end{figure}

\vskip 2mm
$\bullet$ {At least one truncated light-sheet $\TL _k (\BB)$ is future directed, ingoing, and punctured only by black holes.}
\vskip 2mm

\noindent Here, we can make essentially the same argument as in the very first case, since the only difference would be accounting for the black hole entropy (see Fig. \ref{Chimney}). To obtain the same result, we have to use the strengthened form \cite{FMW} of Bousso's bound, mentioned at the end of Section \ref{BoussosHolographicBound}. When we add the black hole entropies and the bound on the entropy passing through  $\TL _k (\BB)$, we get $A/4$. Therefore, the conjecture seems to be correct also in this case.

\vskip 2mm
$\bullet$ {All ({\it i.e.} both) truncated light-sheets $\TL _j (\BB)$ are future directed and punctured by the final singularity, as well as possibly by some black holes.}
\vskip 2mm

\noindent In this case, the surface $\BB$ is not observable and we should not even have listed this possibility here. 

\subsection{Speculative interpretation: Holographic virtual reality}
\label{VirtualReality}

If we think of the second subsystem in the conjecture as a measuring apparatus, then the conjecture gives us a bound on the effective number of states we can distinguish with that apparatus. We know that in quantum mechanics it does not make any sense to describe a physical system with a Hilbert space whose dimension exceeds the number of states we can {\it in principle  } distinguish. If such a Hilbert space was necessary for the description of some system, there would be no limit of an ideal apparatus and the meaning of the concept of quantum states would be completely unclear. Therefore, we might interpret the conjecture as that 
the maximum possible entanglement entropy is equal to the logarithm of the dimension of the Hilbert space which describes the first subsystem.

Note that usually people think that the maximum coarse-grained entropy is directly related to the dimension of the Hilbert space. This assumption is usually correct. Let us consider for example a lattice of spins in non-relativistic quantum mechanics. In general, if the spins are correlated, the fine-grained entropy is not equal to the coarse-grained entropy. However, we can always flip the spins and remove the correlations to obtain a mixed state whose fine-grained entropy is equal to the coarse-grained entropy and also to the logarithm of the dimension of the Hilbert space. Therefore, the maximum coarse-grained entropy is really equal to the logarithm of the dimension of the Hilbert space. The case we are discussing in this paper is completely different. If we assume no volume-extensive entropy at the beginning of time, there is {\it in principle} no way to remove the correlations between the particles produced during reheating and afterwards. The maximum fine-grained entropy (which should be by definition the logarithm of the dimension of the Hilbert space) no longer coincides with the maximum coarse-grained entropy. This might be the overcounting holographic principle talks about.

If the conjecture and this interpretation are correct, there are very interesting consequences when applied to the case of a small region containing the observer with measuring apparatuses which are used to study the rest of the universe. It is no surprise that the entropy of the small region is bounded by the surface area. However, this quantity bounds also the number of different possible results of the observations done by the observer. According to our interpretation of the conjecture, this is precisely the dimension of the Hilbert space which should be used by the observer to describe the rest of the universe. Here we should stress that by `description' of a physical system in quantum mechanics, we always mean nothing else but the description of possible results of our experiments. With this in mind, we can say that in spite of the huge amount of coarse-grained entropy, the number of possible states of the universe can be very small, depending on the size of the region occupied by the observer. This strongly contadicts our every-day concept of objective reality. However, we should not be so much surprised, since it breaks down already on the level of quantum field theory. Let us discuss this issue a little bit closer.

Consider a macroscopic object (for example a pet of a famous physicist),
which is initially in a pure state and which can evolve into two very distinct macroscopic states after some time. Suppose that the `choice' of the final state depends on some quantum-mechanical process. Because of unitarity of the evolution, the whole system will remain in a pure state. However, when we observe the object after a while, we will find it in one of the possible macroscopic states and never in a quantum superposition of them. How is this possible? The answer is easy: During the evolution, the object emits photons and other particles, which are correlated with the subsequent state of the object. The whole system remains  in a pure state, but now, it consists not only of the object itself, but also of the particles emitted. The reduced density matrix of the object itself now corresponds to a mixed state, which we naturally expect. 
This mechanism gives rise to what we perceive as reality, {\it i.e.} it makes things `really happen' instead of being in some quantum superposition of different possibilities.

Let us illustrate the discrepancy between different concepts of reality of different observers on a not too scientific example.  Consider an observer 2060 light-years from us, who has just started 
{\it its}\footnote{We do not specify if the observer is male or female, since these terms have been defined only for some organisms with terrestrial origin.}
long-term program of observing the Earth. Because the photons from the year 44 B.C. still have not reached it, there is a non-zero probability for the observer to see that Julius Caesar was warned against the attempt on his life, defeated the conspirators and had two more children with Cleopatra. However, it does not make any sense for us to ask such a question. We live in a different reality, because the state of things around us is already influenced by the history we know. To be more precise, it corresponds just to a small fraction of the many possible 
states of the photons going to the observer. If we decided to ask the observer what it saw, it would definitely say that Caesar was murdered on March 15, 44 B.C., if it is honest. 

We see that the concept of reality is observer-dependent already in quantum field theory. When we take general relativity into account, the lack of objective reality becomes even more striking. An observer falling into a black hole can cross the horizon, perform physical experiments and finally reach the singularity. On the other hand for an outside observer, no spacetime points under the horizon exist. If there is enough time, the mass of the black hole will be finally converted into Hawking radiation above the horizon and the whole black hole will disappear. 
For the outside observer, the history the infalling observer experienced inside the black hole did not actually happen in the above sense. If it did, the outside observer would have to detect particles going from inside of the black hole, which is clearly impossible. Thus, no objective reality exists.

The interpretation of the conjecture, if correct, tells us about an attractive possibility of `describing all of nature'  by a relatively small Hilbert space from the point of view of any particular observer. It is not important if the observer falls into a black hole or not. Such a description would deal only with the physics relevant to the observer of interest, and it would be an explicit realization of the black hole complementarity principle \cite{BHCP}. In this approach, the observer would be surrounded by a holographic screen representing the rest of the universe, and nothing else would exist. This would resemble virtual reality from the world of computer technology. However, our holographic virtual reality would be so advanced that it could easily compete with the one in \cite{TheMatrix}.

\subsection{A weaker form of the conjecture}
\label{AWeakerFormOfTheConjecture}
The conjecture we have formulated relies on the assumption that `at the beginning of time,' there was no volume-extensive entropy in the universe. Note however, that if we formulated its weaker form where we would restrict ourselves just to regions inside some `causal diamond' as in \cite{NBound}, the corresponding entanglement entropy bound would follow from Bousso's bound almost immediately, without any special assumptions about the early cosmology.  (By a causal diamond we mean a spacetime region which is in the causal past of some event as well as in the causal future of some other event.) Its interpretation in terms of the number of  quantum states would be nevertheless more difficult.

\section{Conclusion}

We have proposed a holographic bound on entanglement entropy inside (or outside) closed space-like surfaces. Also, we have discussed a speculative interpretation of it in terms of holographic properties of a hypothetical theory of quantum gravity. If these thoughts are correct, they  support our hopes that it will be possible to describe nature by a theory which would be directly or indirectly related to superstrings, since string theory is known to incorporate holography at least in certain examples.

\section{Acknowledgements}

I have benefited from useful discussions with Raphael Bousso, Simeon Hellerman, Petr Ho\v rava, Shamit Kachru, Andrei Linde, Liam McAllister,  Lubo\v s Motl, Steve Shenker, Lenny Susskind, Nick Toumbas and Herman Verlinde.
This work was supported in part by an undergraduate fellowship at Charles University, by the Stanford Graduate Fellowship, and by the Institute of Physics, Academy of Sciences of the Czech Republic under Grant No. GA-AV \v CR A1010711.

\end{document}